\documentclass[fleqn,twoside]{article}

\usepackage{espcrc2}
\usepackage{graphicx}
\usepackage{amssymb}

\newcommand{\bc}{\begin{center}}
\newcommand{\ec}{\end{center}}
\newcommand{\be}{\begin{equation}}
\newcommand{\ee}{\end{equation}}
\newcommand{\bea}{\begin{eqnarray}}
\newcommand{\eea}{\end{eqnarray}}
\newcommand{\bi}{\begin{itemize}}
\newcommand{\ei}{\end{itemize}}
\newcommand{\bd}[1]{\begin{dinglist}{#1}}
\newcommand{\ed}{\end{dinglist}}
\newcommand{\bt}{\begin{tabular}}
\newcommand{\et}{\end{tabular}}

\def\ape{\textsf{APE}}
\def\anext{\textsf{apeNEXT}}

\title{%
       \vspace{-7mm}					
       Status of the \anext{} project%
      \thanks{Talk given by D.~Pleiter at the Lattice Conference 2002,
              Cambridge (MA), USA.}
}

\author{%
  \anext{} Collaboration:
  R.~Ammendola\address[roma2]{Physics Department, University of Roma
		``Tor Vergata'' and INFN, Sezione di Roma II, Italy},
  F.~Bodin\address[irisa]{IRISA/INRIA, Campus Universit\'{e} de Beaulieu, Rennes, France},
  Ph.~Boucaud\address[orsay]{LPT, University of Paris Sud, Orsay, France},
  N.~Cabibbo\address[roma1]{INFN, Sezione di Roma, Italy},
  F.~Di~Carlo\addressmark[roma1],
  R.~De Pietri\address[parma]{Physics Department, University of Parma and
		INFN, Gruppo Collegato di Parma, Italy},
  F.~Di~Renzo\addressmark[parma],
  W.~Errico\address[pisa]{INFN, Sezione di Pisa, Italy},
  A.~Fucci\address[CERN]{CERN, Geneva, Switzerland},
  M.~Guagnelli\addressmark[roma2],
  H.~Kaldass\address[ztn]{DESY Zeuthen, Germany},
  A.~Lonardo\addressmark[roma1],
  S.~de~Luca\addressmark[roma1],
  J.~Micheli\addressmark[orsay],
  V.~Morenas\address[clermont]{LPC, Universit\'{e} Blaise Pascal and IN2P3,
		Clermont, France}
  O.~Pene\addressmark[orsay],
  R.~Petronzio\addressmark[roma2],
  F.~Palombi\addressmark[roma2],
  D.~Pleiter\address[NIC]{NIC/DESY Zeuthen, Germany},
  N.~Paschedag\addressmark[ztn],
  F.~Rapuano\addressmark[roma1],
  P.~De~Riso\addressmark[roma2],
  D.~Rossetti\addressmark[roma1],
  A.~Salamon\addressmark[roma2],
  G.~Salina\addressmark[roma2],
  L.~Sartori\addressmark[pisa],
  F.~Schifano\addressmark[pisa],
  H.~Simma\addressmark[ztn],
  R.~Tripiccione\address[ferrara]{Physics Department, University of Ferrara, Italy}
  P.~Vicini\addressmark[roma1]
}

\begin{document}

\begin{abstract}
\vspace*{-0.2cm}
We present the current status of the \anext{} project. Aim of this project
is the development of the next generation of \ape{} machines which will
provide multi-teraflop computing power. Like previous machines, \anext{} is
based on a custom designed processor, which is specifically optimized
for simulating QCD. We discuss the machine design, report
on benchmarks, and give an overview on the status of the software
development.
\vspace*{-0.5cm}
\end{abstract}

\maketitle

\section{INTRODUCTION}

The \anext{} project was initiated \cite{proposal} with the goal to
build supercomputers with a peak performance of more than 5 TFlops
and a sustained efficiency of $O(50\%)$ for key lattice gauge theory
kernels. Aiming for both large scale simulations with dynamical fermions
and quenched calculations on very large lattices the architecture should
allow for large on-line data storage and input/output channels to sustain
$O(0.5)$ MByte per second per GFlops.  Finally, the programming environment
should allow smooth migration from older APE systems, i.e.~support the
TAO language, and provide the C language with comparable performance.

Although there are a number of similarities between the architecture of
\anext{} and former generations of \ape{} supercomputers, there were a
number of design challenges to be solved in order to meet the machine
specifications outlined above. For \anext{} all processor functionalities,
including the network devices, were to be integrated into one single custom
chip running at a clock frequency of 200 MHz. Unlike former machines,
the nodes will run asynchronously, which means that \anext{} follows
the single program multiple data (SPMD) programming model.

\vspace*{-0.2cm}
\section{PROCESSOR AND GLOBAL DESIGN}

\begin{figure}[t]
\bc
\includegraphics[scale=0.31]{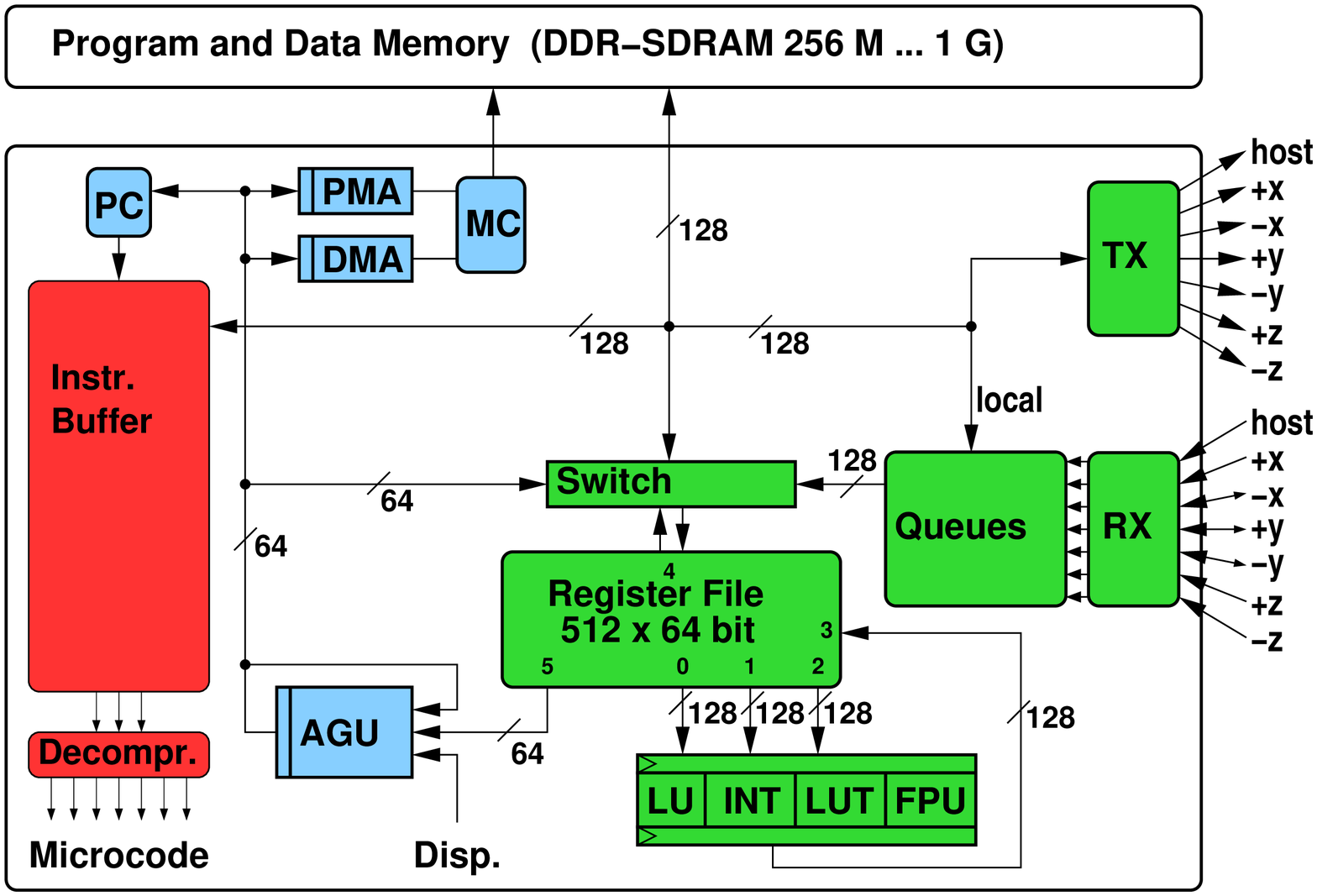}
\vspace*{-1.5cm}
\ec
\caption{The \anext{} processor.\label{fig:processor}}
\vspace*{-0.8cm}
\end{figure}

The \anext{} processor is a 64-bit architecture. Its arithmetic unit
can at each clock cycle perform the \ape{} normal operation $a\times b+c$,
where $a$, $b$, and $c$ are IEEE double precision complex numbers.
The peak performance of each node is therefore 1.6 GFlops. Like
previous \ape{} computers \anext{} provides a very large register
file of 256 (64+64)-bit registers. Selected details of the processor
are shown in Fig.~\ref{fig:processor}.

The memory interface of \anext{} supports DDR-SDRAM from 256 MByte upto
1 GByte.  The memory is used to store both data and program
instructions. Conflicts between data and instruction load-operations
are therefore likely. These could easily become significant
since \anext{} is a microcoded architecture controlled by 128-bit very long
instruction words. Two strategies have been employed to
avoid these conflicts. First, the hardware supports compression of
the microcode. The compression rate usually depends on the level of
optimization, typically it is in the range of 40-70\%. Second, an
instruction buffer allows pre-fetching of (compressed) instructions.
Controlled by software a section of the instruction buffer can be used
to store performance critical kernels for repeated execution.

Each \anext{} node contains seven LVDS link interfaces which allow
for concurrent send and receive operations. Once a communication
request is queued it is executed independently of the rest of the
processor, which is a pre-requisite for overlapping network and
floating point operations. Each link is able to transmit one
byte per clock cycle, i.e.~the gross bandwidth is 200 MByte per second
per link. Due to protocol overhead the effective network bandwidth
is $\le 180$ MByte per second. The network latency is
$O$(0.1~$\mu$s) and therefore at least one order of magnitude smaller
than for today's commercial high performance network
technologies.

\begin{table}[t]
\caption{Key machine parameters}
\bt{|l|l|}
\hline
clock frequency    & 200 MHz \\
peak performance   & 1.6 GFlops \\
memory             & 256-1024 MByte/node \\
memory bandwidth   & 3.2 GByte/sec \\
network bandwidth  & 0.2 GByte/sec/link \\
register file      & 512 registers \\
instruction buffer & 4096 words \\
\hline
\et
\vspace*{-0.8cm}
\end{table}

Six of these link interfaces are used for connecting each node to its
nearest neighbours within a three-dimensional network.
The seventh link of up to one node per board can be used as an
I/O channel by connecting it to an external front-end PC equipped
with a custom PCI-LVDS interface card.  The number of external links
and therefore the total I/O bandwidth can be flexibly adapted
to the needs of the users.
Although all nodes are connected to their nearest neighbours only, the
hardware allows routing across up to three orthogonal links to
all nodes on a cube, i.e.~connecting nodes at distance
$(\Delta_x,\Delta_y,\Delta_z)$ with $|\Delta_i| \le 1$.

Although the network bandwidth is large compared to other network
technologies, it is significantly smaller than the local memory bandwidth.
It is therefore mandatory to support efficient mechanisms for data
pre-fetching. For this purpose a set of pre-fetch queues is provided.
Pre-fetch instructions in a user
program will initiate the memory controller and, in case of remote data,
the network to move the requested data into the queues.
At a later stage of program execution this data is loaded from the queues
into the register file in the same order as the pre-fetch instructions had
been issued. Only if the data is not available
at that point
the processor will be halted until the data has arrived.

\begin{figure}[t]
\bc
\vspace*{0.1cm}
\includegraphics[scale=0.38]{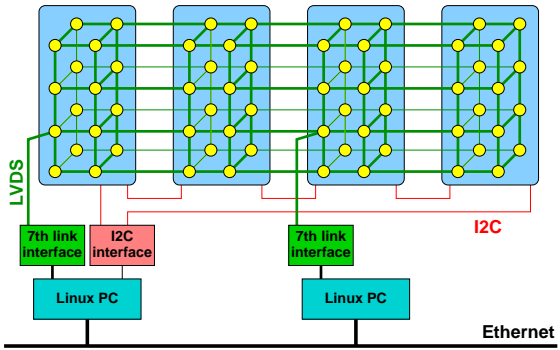}
\vspace*{-1.5cm}
\ec
\caption{Possible \anext{} configuration with 4 boards, 2 external
LVDS-links for I/O, and a chained I2C-link for slow-control.\label{fig:global}}
\vspace*{-0.5cm}
\end{figure}

The global design of \anext{} is shown in Fig.~\ref{fig:global}. There
will be 16 \anext{} processors on one processing board and 16 boards
will be attached to one backplane. Each node is connected to a simple
I2C-link used for bootstrapping and controlling the machine.

\section{SOFTWARE AND BENCHMARKS}

We will provide both a TAO and a C compiler for \anext{}. The latter
is based on the freely available {\em lcc} compiler \cite{lcc} and
supports most of the ANSI 89 standard with a few language extensions
required for a parallel machine.
For machine specific optimizations at the assembly level,
e.g.~address arithmetics and register move operations,
the software package {\em sofan} is under development.
Finally, the
microcode generator ({\em shaker\/}) optimizes instruction scheduling,
which for \ape{} machines is completely done by software.

For all parts of the compiler software stable prototype versions are
available and were already used to benchmark the \anext{} design.
For this purpose we considered various typical linear algebra operations
like the product of two complex vectors. This operation is basically
limited by the memory bandwidth, implying a maximum
sustained performance of 50\%. From VHDL simulations that include
all machine details the efficiency
was found to be 41\%. Even higher performance rates can be achieved for
operations requiring more floating point operations per memory access,
like multiplying arrays of SU(3) matrices, which achieves
an efficiency of 65\%. In QCD simulations most of
the time is spent applying the Dirac operator, e.g.~the
Wilson-Dirac operator $M=1-\kappa H$.
We therefore investigated the operation $H \psi$ for which a sustained
performance of 59\% has been measured. This figure is made possible by
extensive use of the pre-fetch features of the processor, and keeping a
local copy of the gauge fields to save network bandwidth. Even for the
smallest local lattices complete overlap of floating point operations and
network communication is possible, so the time when the processor waits
for data is close to zero.

\section{\anext{} PC PROJECT}

While pursuing the aim of building a custom designed multi-teraflop
computer the \ape{}-collaboration started activities to develop a
fast network, which is also based on LVDS, for interconnecting PCs.
The final network interface is planned to consist of two bi-directional
links with a bandwidth of 400 MByte per second each. Presently,
a test setup with two PCs is running stable using prototype interfaces
with one link each and a bandwidth of 180 MByte per second. For
this setup running a QCD solver code the sustained network bandwidth
was found to be 77 MByte per second. A similar setup using the final
network interfaces is expected to come into operation in September 2002.

\section{OUTLOOK AND CONCLUSIONS}

The hardware design of the next generation of \ape{} custom built computers
has been completed. While prototype boards and backplane are available
since the end of 2001, a prototype \anext{} processor is expected to
be ready by the end of 2002. A larger prototype installation is planned to
be running by middle of 2003. There exists a stable prototype version for
all parts of the compiler software. Based on this software we were
able to demonstrate that key lattice gauge theory operations will be
able to run at a sustained performance of $O(50\%)$ or more.


\end{document}